\documentstyle[art12]{article}
\newcommand\sect[1]{\setcounter{equation} 0\section{#1}}

\newcommand{\be}{\begin{equation}}
\newcommand{\ee}{\end{equation}}
\newcommand{\ba}{\begin{eqnarray}}
\newcommand{\ea}{\end{eqnarray}}
\newcommand{\baa}{\begin{eqnarray*}}
\newcommand{\eaa}{\end{eqnarray*}}
\newcommand{\bb}{\begin{thebibliography}}
\newcommand{\eb}{\end{thebibliography}}
\newcommand{\ci}[1]{\cite{#1}}
\newcommand{\bi}[1]{\bibitem{#1}}
\newcommand{\lab}[1]{\label{#1}} 
\newcommand{\re}[1]{(\ref{#1})}
\newcommand{\Tr}{\mbox{Tr\,}}

\newcommand\e{\mbox{e}}
\newcommand\CO{{\cal O}}
\newcommand\geff{\bar g}
\newcommand\ro[1]{\sqrt{(#1-\lambda_+)(#1-\lambda_-)}}
\newcommand\sq[1]{\sqrt{{#1}^2+c}}
\newcommand\ra[1]{\sqrt{1+\frac{c}{{#1}^2}}}
\newcommand\qqquad{\qquad\quad}
\newcommand\qqqquad{\qquad\qquad}
\newcommand\fac[2]{\mbox{$\frac{#1}{#2}$}}
\newcommand\vev[1]{\langle{#1}\rangle}
\newcommand\VEV[1]{\left\langle{#1}\right\rangle}

\begin{document}
\thispagestyle{empty}
\hfill {\sc UPRF--92--334} \qquad { } \par
\hfill {\sc hepth@xxx/9205014 } \qquad { } \par
\hfill April, 1992 \qquad { }

\vspace*{15mm}

\begin{center}
\renewcommand{\thefootnote}{\fnsymbol{footnote}}
{\LARGE Matrix Model Perturbed by Higher Order
 \\[4mm]  Curvature Terms}

\vspace{12mm}

\newcommand{\st}{\fnsymbol{footnote}}%

{\large G.~P.~Korchemsky}%
\footnote{On leave from the Laboratory of Theoretical Physics,
          JINR, Dubna, Russia}
\footnote{INFN Fellow}

\medskip

{\em Dipartimento di Fisica, Universit\`a di Parma and \par
INFN, Gruppo Collegato di Parma, I--43100 Parma, Italy} \par
{\tt e-mail: korchemsky@vaxpr.cineca.it}

\end{center}

\vspace*{20mm}

\begin{abstract}
The critical behaviour of the $D=0$ matrix model with potential
perturbed by nonlocal term generating touchings between random
surfaces is studied. It is found that the phase diagram of the
model has many features of the phase diagram of discretized
Polyakov's bosonic string with higher order curvature terms
included. It contains the phase of smooth (Liouville) surfaces,
the intermediate phase and the phase of branched polymers.
The perturbation becomes irrelevant at the first phase and
dominates at the third one.
\end{abstract}

\newpage

\renewcommand{\thefootnote}{\arabic{footnote}}
\setcounter{footnote}{0}

\sect{Introduction}

After a recent progress, initiated by KPZ paper \ci{KPZ}, in the investigation
of conformal invariant field models with central charge $c\leq 1$ coupled to
gravity, an understanding of physics beyond the ``barrier'' $c=1$ is still
an open question. In string theory, $c >1,$ KPZ approach leads to complex
anomalous dimensions of fields indicating that an instability is developed
in the theory.

The same phenomena was noticed independently under computer
simulations of discretized Polyakov's bosonic string. It was found \ci{Num}
that near the point with $c=1$ transition to the phase of crumpled surfaces
occurs. In this phase the world sheet of string has a form of ``hedgehogs''
and an underlying effective theory cannot be a string theory. One of the
well known ways to cure this instability is to perturb the string action by
adding higher order intrinsic curvature terms, like $\int d^2\sigma \sqrt{g}
R^2.$ It turned out \ci{Num} that the discretized perturbed model has four
different phases depending on the values of central charge and
constant coupled to the intrinsic curvature: phases of crumpled surfaces,
branched polymers, intermediate phase and phase of smooth (Liouville)
surfaces. It is in the last two phases where one may hope to construct an
effective string theory with the central charge $-\infty < c < 4.$ For
$-\infty < c \leq 1$ the model has a remarkable universality property:
additional curvature terms become irrelevant in the continuum limit and the
critical behaviour is governed only by the central charge, in accordance with
KPZ relations leading to negative values of string susceptibility exponent
$\gamma_{str} <0.$ At the same time in the intermediate phase, for
$1 < c < 4,$ the curvature terms are responsible for nontrivial critical
behaviour of the model with positive string susceptibility exponent
\be
      0 < \gamma_{str} < 0.5
\lab{1.1}
\ee
An understanding of this phase in the continuum theory is still lacking
although possible mechanism of its appearance was advocated in \ci{Pol}.
It was proposed \ci{Touching,g>0} to use the matrix models to justify the
critical behaviour \re{1.1}.

Matrix models allow us to reproduce KPZ results for conformal matter with
$c \leq 1$ \ci{Review}. In order to get the critical behaviour \re{1.1} it
seems natural to perturb the potential of the matrix models by terms which
include the effects of higher order curvature terms. Possible form of such
terms was discussed in \ci{Touching}. The simplest one was found to be
$(\Tr M^2)^2$ and possible matrix model, $D=0$ hermitian one-matrix model
with the potential $\Tr(M^2 + g M^4)$ perturbed by this term, was considered
in \ci{Touching}. The appearance of additional phase with positive exponent
$\gamma_{str}=1/3$ was found and the existence of a series of analogous
phases with $\gamma_{str}=1/n,$ $(n=3,\ 4,\ldots)$ was conjectured. Although
these points lie in the region \re{1.1} their interpretation as critical
points of $c > 1$ matter coupled to gravity remains unclear.

In the present paper we investigate the influence of the perturbation
$(\Tr M^2)^2$ on the critical behaviour of $D=0$ hermitian one-matrix model
with an arbitrary polynomial potential. We identify the critical points of
the model and find that the resulting phase diagram has many features of the
phase diagram of the discretized Polyakov's string with higher order
curvature terms included \ci{Num}. Under increasing of the constant coupled to
perturbation the system passes through the phase of smooth surfaces,
intermediate phase and the phase of branched polymers. It turns out that
the perturbation is irrelevant at the first phase and dominates at the
last one. We show that in the continuum limit the intermediate phase with
positive exponent $\gamma_{str}=1/(K+1)$ $(K\geq 2)$ appears only as
perturbation of $(2,2K-1)$ minimal conformal model coupled to gravity.

\sect{Potential of perturbed model}

The partition function of $D=0$ hermitian matrix model is defined as
\be
\e^{Z(\alpha)} = \int dM\, \exp(-\alpha N \Tr V(M))
\lab{2.1}
\ee
where $\alpha$ is the bare cosmological constant, $V(M)$ is the potential of
the model and integration is performed over hermitian $N\times N$ matrices
$M.$ For large $N$ and $\alpha$ close to the critical value $\alpha_{cr},$
at which the perturbation series for $Z(\alpha)$ become divergent, the string
susceptibility has a behaviour
\be
\chi=\frac1{N^2} \frac{d^2 Z(\alpha)}{d\alpha^2}
     \sim (\alpha - \alpha_{cr})^{-\gamma_{str}}
\lab{2.2}
\ee
with nontrivial critical exponent $\gamma_{str}.$ It is well known
\ci{Review}, that in $D=0$ hermitian one-matrix model the string
susceptibility exponent can have two different kinds of values associated
with multicritical Kazakov models \ci{Multi} and Penner model
\ci{Penner1,Penner2}. In the $K-$th multicritical model the potential has
the following form
\be
V(M)\equiv U_{2K}(M)=\sum_{n=1}^{K} (-1)^{n-1}
                     \frac{(n-1)! K!}{(2n)! (K-n)!}M^{2n}
\lab{2.3}
\ee
and the exponent $\gamma_{str}$ takes negative values
\be
\gamma_{str}=-\frac{1}{K}.
\lab{2.4}
\ee
This model was identified in the continuum limit as $(2,2K-1)$ minimal
conformal model coupled to 2D gravity \ci{MinCFT}. In the Penner model
with nonpolynomial potential
\be
V(M)=\log(1-M)+M
\lab{2.5}
\ee
the critical exponent is equal to
\be
\gamma_{str}=0
\lab{2.6}
\ee
and in the continuum limit model possesses many properties of $c=1$
quantum gravity \ci{Penner1,Penner2}. In both cases \re{2.4} and \re{2.6}
the exponents $\gamma_{str}$ are in agreement with KPZ predictions.

Let us perturb the potentials \re{2.3} and \re{2.5} to include higher
order curvature terms. The simplest way to do it is to add the following
``nonlocal'' interaction term $(\Tr M^2)^2$ to the potential \ci{Touching}.
To this end one chooses the potential of perturbed matrix model as
\be
\Tr V(M) = \Tr V_0(M) - \frac{g\alpha}{4N} (\Tr M^2)^2
\lab{2.7}
\ee
where $V_0(M)$ is the generalized Penner potential%
\footnote{The matrix model with potential $V_0(M)$ was considered in
          \ci{GenPen} where its nontrivial critical behaviour was found
          provided that one identifies cosmological constant with the
          dimension of the matrix $M.$}
\be
\Tr V_0(M) = \int dx\, \rho(x) \Tr \log(x-M)
\lab{2.7.1}
\ee
and $\rho(x)$ is the normalized function with well defined moments
$$
\int dx\,\rho(x)=1 \qqqquad \int dx\,x^{-n}\rho(x)=\mbox{finite}
$$
having a meaning of the density of external field. The reason for this
particular choice of the potential is twofold. First of all, expanding
potential $V_0(M)$ in powers of $M$ it is possible to get an arbitrary
polynomial potential by tunning of the moments $\int dx\,x^{-n}\rho(x).$
Secondly, it turns out that the presence of an external field $\rho(x)$
simplifies the solution of the model.

To understand the meaning of the perturbation $(\Tr M^2)^2$ one considers
the form of random surfaces generated by the potential \re{2.7}. The first
term, $V_0(M),$ being expanded in powers of $M$ leads in a standard way to
a sum over discretized (in general unconnected) closed surfaces of an arbitrary
genus. The last term in \re{2.7} opens a possibility for these surfaces to
touch each other. Then, unconnected surfaces with touchings included become
connected and have a form of ``soap bubbles'' \ci{Touching}.  Note, that
the touchings may change the genus of the surface.

In what follows we consider the model in the leading $N \to \infty$
(spherical) approximation which allows us to identify the possible critical
points. In the spherical approximation any two planar surfaces may touch
each other only at once and cannot form closed chains of soap bubbles. A
power counting \ci{Review} indicates that the partition function \re{2.1}
has the following decomposition for large $N$
\be
Z(\alpha) = N^2 \sum_{S,T} Z(S,T) \alpha^{-S} g^T + \CO(N^0)
\lab{2.8}
\ee
where $S$ is the total area of the discretized planar surface, $T$ is the
number of touchings this surface has and $Z(S,T)$ is the partition function
for fixed $S$ and $T.$ This expression allows us to identify the bare
cosmological constant as $\alpha=\e^{\Lambda_B} >0.$ In the continuum limit
when $\alpha$ approaches the critical value $\alpha_{cr}$ the surfaces with
area $S\to\infty$ dominate in the sum \re{2.8} with the following asymptotics
$
\sum_T Z(S,T) g^T \sim \alpha_{cr}^S(g) S^{\gamma_{str}(g)-3},
$
leading to the critical behaviour of the partition function
$
Z(\alpha) \sim (\alpha-\alpha_{cr}(g))^{2-\gamma_{str}(g)}.
$
Here, the exponent $\gamma_{str}(g)$ and the critical value of cosmological
constant are functions of touching coupling constant. Formally,
$\alpha_{cr}(g)$ is given by series in $g$ and nonanalyticity of this
function for some $g$ would mean that the critical behaviour of the model
is determined by the surfaces with infinite number of touchings.
One expects that in the limit $\alpha_{cr}(g) \gg g$ when it is more
favorable for the critical surfaces to form a touching instead of increase
area the critical surfaces are degenerated into infinite chains of ``soap
bubbles'' and the model turns into the phase of branched polymers
\ci{Touching}.

To find the phase diagram of the model in $(\alpha, g)$ plane we calculate
the string susceptibility. Using the definition \re{2.2} and an
explicit form of the potential one gets
$$
\chi=-\frac{d{\ }}{d\alpha}\left(\int dx\,\rho(x)
      \VEV{\frac1{N}\Tr\log(x-M)}
         -\frac{\alpha g}2\VEV{\left(\frac1{N}\Tr M^2\right)^2}
         \right)
$$
where for an arbitrary operator
$\vev{A(M)} \equiv \e^{-Z(\alpha)}\int dM\,A(M)\exp(-\alpha N\Tr V(M))$
denotes the connected correlator. Here, the second term is responsible for the
touchings of random surfaces. It can be simplified for large $N$
using the factorization property of the vacuum averaged product of invariant
operators \ci{Loop1}:
\be
\langle (\Tr M^2)^2\rangle =(\langle \Tr M^2 \rangle)^2 + \CO(N^{0})
\lab{2.9}
\ee
This property implies that in the spherical approximation the touchings of the
surfaces are measured effectively with the following constant
\be
\geff=-g\alpha \VEV{\frac1{N}\Tr M^2}
\lab{2.10}
\ee
called effective touching coupling constant. Then, for the string
susceptibility we get
$$
\chi=-\frac{d{\ }}{d\alpha}\left(\int dx\,\rho(x)
      \VEV{\frac1{N}\Tr\log(x-M)}-\frac{\geff^2}{2g\alpha}\right)
$$
The explicit form of this expression suggests us to connect $\chi$ with one
loop correlator
$$
\vev{W(z)} = \VEV{\frac1{N}\Tr\frac1{z-M}},
$$
using a simple identity
$$
\int_{\Lambda}^{x} dz\, \vev{W(z)} = \VEV{\frac1{N}\Tr\log(x-M)} - \log\Lambda,
\qquad \mbox{as $\Lambda\to\infty$}
$$
Here, the divergence of the integral as $\Lambda\to\infty$ is a consequence
of the asymptotics of loop correlator
\be
\vev{W(z)} \to 1/z, \qquad \mbox{as $z\to\infty.$}
\lab{2.11}
\ee
Finally, the expression for string susceptibility is given by
\be
\chi=-\frac{d{\ }}{d\alpha}\left(\int dx\,\rho(x)
     \int_{\Lambda}^{x} dz\, \vev{W(z)}
         +\log\Lambda -\frac{\geff^2}{2g\alpha}
         \right).
\lab{2.12}
\ee
One completes this expression by the relation between effective touching
coupling constant \re{2.10} and one loop correlator
\be
\geff = -g\alpha \oint_C \frac{dz}{2\pi i}\ z^2 \vev{W(z)}
\lab{2.13}
\ee
where integration contour $C$ encircles singularities of $\vev{W(z)}.$
Equations \re{2.12} and \re{2.13} allow us to find the string
susceptibility starting from one loop correlators.

\sect{Perturbed loop equation}

To calculate loop correlators one uses a powerful method of loop equations
\ci{Loop1,Multi,Loop2,Loop3}. In $D=0$ hermitian matrix model with potential
$U(M)$ the solution of loop equations gives us (one-cut) loop correlator in
the limit $N\to\infty$ as \ci{Loop3}
\be
\vev{W(z)} = \frac{\alpha}2 \oint_C \frac{d\omega}{2\pi i}
 \frac{U^\prime(\omega)}{z-\omega}\frac{\ro{z}}{\ro{\omega}}
\lab{3.1}
\ee
where boundaries $\lambda_\pm$ of the cut are fixed by the asymptotics
\re{2.11}. Since the potential of the model \re{2.7} contains an additional
nonlocal term the modification of the loop equation is needed. The perturbed
loop equation we get in a standard way \ci{Loop2} performing the change of
the integration variables $M\to M+\frac{\varepsilon}{z-M}$ in \re{2.1} with
infinitesimal $\varepsilon$
$$
\vev{W(z) W(z)} =\alpha \oint_C\frac{d\omega}{2\pi i}\frac1{z-\omega}
\VEV{(V_0^\prime(\omega)-\omega\frac{g\alpha}{N}\Tr M^2)W(\omega)}
$$
Using relation analogous to \re{2.13} it is easy to express $\Tr M^2$ as an
integral of $W(z^\prime)$ and then to get a closed equation for loop
correlators. In the limit $N\to\infty$ we apply factorization property
\re{2.9} to transform the loop equation into
\be
\vev{W(z)}^2 = \alpha\oint_C\frac{d\omega}{2\pi i}
               \frac{U^\prime(\omega)}{z-\omega}\vev{W(\omega)}
\lab{3.1.1}
\ee
where the notation was introduced
\be
U^\prime(z) = V_0^\prime(z) - z g \alpha \VEV{\frac1{N}\Tr M^2}
       \equiv -\int \frac{dx \rho(x)}{x-z} + z \geff
\lab{3.2}
\ee
The solution of the equation \re{3.1.1} is given by \re{3.1} provided that
one considers $U^\prime(z)$ as a derivative of some potential. Let us choose
for simplicity the density $\rho(x)$ to be an even function
$$
\rho(-x)=\rho(x).
$$
Then, the potential $V(z)$ is also even and, as a consequence, the cut of the
loop correlator is symmetric \ci{Loop3} with $\lambda_+=-\lambda_-.$
After substitution of \re{3.2} into \re{3.1} the loop correlator is equal to
\be
\vev{W(z)} = \frac{\alpha}2\int \frac{dx\,\rho(x)}{z-x}
\left(1-\frac{\sq{z}}{\sq{x}}\right)+\frac{\alpha}2\geff (z-\sq{z})
\lab{3.3}
\ee
and the additional condition \re{2.11} leads to the following equation for
$c=\lambda_+\lambda_-=-\lambda_+^2 < 0$
\be
\geff = 2 f(c)-\frac4{\alpha c}
\lab{3.4}
\ee
Here, the notation was introduced for the function
\be
f(c)=\int\frac{dx\,\rho(x)}{\sq{x}\ (x+\sq{x})}
    =\frac1{c}\int\,dx\rho(x)\left(1-\frac1{\ra{x}}\right)
\lab{3.5}
\ee
depending on the moments of density $\rho(x).$ For latter convenience
one defines the following functions
$$
\psi(c) = \int\frac{dx\,\rho(x)}{(x+\sq{x})^2},
\qqqquad
\phi(c) = c \int\frac{dx\,\rho(x)}{\sq{x}\ (x+\sq{x})^3}
$$
They are related to each other by simple equations
\be
f^\prime(c) + \phi^\prime(c) = -\frac{2}{c} \phi(c),
\qqqquad
f(c) + \phi(c) = 2 \psi(c)
\lab{3.6}
\ee
or equivalently
\be
\phi(c)=-\frac{1}{c^2}\int_{0}^{c} dt\,t^2 f^\prime(t),
\qqqquad
\psi(c)=\frac{1}{c^2}\int_{0}^{c} dt\,t f(t)
\lab{3.7}
\ee
Note, that loop correlator \re{3.3} depends on the effective touching coupling
constant which can be found in its turn after substitution of $\vev{W(z)}$
into \re{2.13} as
\be
\geff=\frac{4g(\alpha c)^2}{16+g(\alpha c)^2}
      \int\frac{dx\,\rho(x)x}{\sq{x}\ (x+\sq{x})^2}
     =\frac{2g(\alpha c)^2}{16+g(\alpha c)^2}(f(c)-\phi(c))
\lab{3.8}
\ee
Being combined together equations \re{3.3}, \re{3.4} and \re{3.8} define
one loop correlator in the spherical approximation.

\sect{String susceptibility}

To evaluate the string susceptibility we substitute one loop correlator
\re{3.3} into relation \re{2.12}. After integration one gets in the limit
$\Lambda\to\infty$
$$
\int dx\,\rho(x)\int_{\Lambda}^{x} dz\,\vev{W(z)}+\log\Lambda
=
\frac{\alpha}{2}\int dx\int dy\,\rho(x)\rho(y)
                                 \log\left(\frac{xy+c+\sq{x}\sq{y}}2\right)
$$
\be
+(1-\alpha)\int dx\,\rho(x)\log\left(\frac{x+\sq{x}}2\right)
+\frac{\alpha}4\geff\int dx\,\rho(x)x(x-\sq{x})+\frac{\alpha}8\geff c
\lab{4.1}
\ee
As follows from \re{2.12}, the string susceptibility contains a full
derivative of this expression with respect to $\alpha.$ Its calculation is not
trivial because both $c$ and $\geff$ depend on $\alpha.$ However, expression
\re{4.1} has a remarkable property: its partial derivative with respect to $c$
equals zero as only one takes into account condition \re{3.4}. Hence,
differentiating \re{4.1} we may ignore the dependence of $c$ on $\alpha$ and
after some transformations
\ba
\chi&=&-\frac12\int dx\int dy\,\rho(x)\rho(y)
     \log\left(2\frac{\ra{x}+\ra{y}}
                     {\left(1+\ra{x}\right)\left(1+\ra{y}\right)}
         \right)
\nonumber \\
   & & +\frac{\geff}{g\alpha}\frac{d\geff}{d\alpha}
     \left(1-\frac{g(\alpha c)^2\psi(c)}{8\geff}\right)
    -\frac{\geff^2}{2g\alpha^2}
     \left(1+\frac{g(\alpha c)^2\psi(c)}{4\geff}\right)
\lab{4.2}
\ea
Here, $\geff$ is given by \re{3.8} and the equation for $c$ we find from
\re{3.4} and \re{3.8} as
\be
\alpha c = \frac{2(16+g(\alpha c)^2)}{16f(c)+g(\alpha c)^2\phi(c)}
\lab{4.3}
\ee
Solving the system of equations \re{4.2}, \re{4.3} and \re{3.8} one can
evaluate the string susceptibility $\chi$ for an arbitrary even density
$\rho(x).$ The properties of $\chi$ and possible critical behaviour of
the model are considered in the next section.

\sect{Analysis of critical behaviour}

It is convenient to start the analysis of critical behaviour by considering
the special case $g=0$ when the potential \re{2.7} reduces to the potential
\re{2.7.1} of so called ``reduced'' hermitian matrix model with even density
$\rho(x).$

\subsection{Unperturbed critical behaviour}

For $g=0$ the equations \re{4.3} and \re{3.8} are replaced by
\be
\frac{2}{\alpha} =c f(c) \qqqquad \geff = 0
\lab{5.1}
\ee
The string susceptibility is given by the first term in \re{4.2} and depends
on $\alpha$ only through the function $c=c(\alpha).$ Therefore, there appear
the following two sources of nonanalyticity of the string susceptibility
$\chi=\chi(c(\alpha)):$ singularities in the dependence of $\chi$ on $c$
(Penner's point) or in the dependence of $c$ on $\alpha$ (Kazakov's
multicritical points). For instance, for even density $\rho(x)$ having a
form
\be
\rho(x)=\fac12(\delta(x-\eta)+\delta(x+\eta))
\lab{5.2}
\ee
the string susceptibility \re{4.2} has a logarithmic singularity at
$c_{cr}=-\eta^2.$ The corresponding critical value of the cosmological
constant can be found from \re{5.1}, \re{5.2} and \re{3.5} to be
$\alpha_{cr}=\frac2{c_{cr}f(c_{cr})}=0$ and near this
point $\alpha=-2\ra{\eta}.$ Hence, the string susceptibility \re{4.2} has a
logarithmic behaviour \ci{Penner1,Penner2}
\be
\chi = -\fac12 \log\alpha
\lab{5.3}
\ee
as $\alpha\to 0.$ Such a behaviour of $\chi$ is one of the features of
$c=1$ quantum gravity. Note, that despite of the original Penner model
\ci{Penner1,Penner2} the
critical point $\alpha=1$ did not appear here because the underlying mechanism
of criticality based on the shrinking into point of the boundaries of the
cut of one loop correlator $(\lambda_+\to\lambda_-\neq 0)$ is suppressed for
even potential $V_0(M)$ leading to the symmetric cut with $\lambda_+=
-\lambda_-.$ Penner critical point originates from the singular dependence
of $\chi$ on $c.$ Otherwise, for $\chi$ having a well defined expansion in
powers of $c,$ the nonanalyticity of the function $c=c(\alpha)$ gives rise to
criticality. In particular, by choosing the function $f(c)$ defined in
\re{3.5} as
\be
cf(c) = 2\left(1-\left(1+\frac{c}4\right)^K\right) = \frac2{\alpha}
\lab{5.4}
\ee
one gets $c-c_{cr}=4(\alpha-\alpha_{cr})^{1/K}$ for $\alpha_{cr}=1$
and $c_{cr}=-4.$ Then, the differentiation of \re{4.2} yields
$$
\left.\frac{d\chi}{dc}\right|_{g=0}=\frac14 c f^2(c)
$$
and near the critical point ($\alpha\to\alpha_{cr}$)
the string susceptibility is equal to
$$
\chi=-\fac14(c-c_{cr})=-(\alpha-\alpha_{cr})^{1/K}
$$
leading to the $K-$th multicritical point with the exponent \re{2.4}. By
comparing the explicit form \re{5.4} of the function $f(c)$ with the
definition \re{3.5}, we find the moments of the density $\rho(x)$ and
restore the corresponding potential $V_0(M)$ to be equal to \re{2.3}.

\subsection{Perturbed critical behaviour}

For $g\neq 0$ when touchings of the surfaces become to play the role
the string susceptibility \re{4.2} acquires the additional term depending
on the
effective touching coupling constant $\geff$ and its derivative. As a
result, there appears a new mechanism of criticality additional to that
considered before for $g=0.$ This mechanism explores the singularities in
the dependence of $\frac{d\geff}{d\alpha}$ on $\alpha$ and leads, as it
will be shown below, to the appearance of new critical points with positive
exponents $\gamma_{str}.$

We begin the analysis of the system of equations \re{4.2}, \re{4.3} and
\re{3.8} by calculating
the derivative $\frac{d\alpha}{d c}.$ The zeros of this derivative would
mean the nonanalyticity of the function $c=c(\alpha)$ which is one of the
sources of criticality. Differentiation of the both sides of \re{4.3} gives
\be
\frac{d\alpha}{dc}=-\ \ \frac{\alpha^2c}{2(16-g(\alpha c)^2(1-\alpha
c\phi(c)))}
                    \left(f^\prime(c)+\frac2{\alpha c^2}\right)
                    \left(16-g(\alpha c)^2\right)
\lab{5.5}
\ee
In the special case $g=0$ this expression with \re{5.4} reduces to
$$
\left.\frac{d\alpha}{dc}\right|_{g=0}
\sim \frac{k}4\left(1+\frac{c}4\right)^{K-1},
\qquad \mbox{as $c\to c_{cr}=-4$}
$$
and the derivative has $K-$th order zero. One finds that for $g\neq 0$ the
equation $\frac{d\alpha}{dc}=0$ has two different branches of solutions:
\be
g(\alpha_{cr} c_{cr})^2 =16
\lab{5.6}
\ee
and
\be
f^\prime(c_{cr})+\frac2{\alpha_{cr} c_{cr}^2}=0
\lab{5.7}
\ee
It is interesting to note that the first equation does not depend on the
explicit form of the density $\rho(x),$ whereas the second one does not depend
on the touching coupling constant. Hence, the appearance of the \re{5.6} is a
general property of a matrix model perturbed by touchings interaction
term. It is this solution that leads to the phase of branched polymers for
$g > g_0$ when only touching term is responsible for the formation of
the critical surfaces and an explicit form of the potential $V_0(M)$ turns
out to be unessential. At the same time, only solution \re{5.7} survives in
the limit $g\to 0$ and it corresponds to the phase of smooth surfaces.

As we saw before, in the unperturbed matrix model with potential $V_0(M)$
the existence of solutions of the equation $\frac{d\alpha}{d c}=0$ leads to
negative critical exponents $\gamma_{str}$ for which the string
susceptibility \re{2.2}
is regular as $\alpha\to\alpha_{cr}.$ In our case, $g\neq 0,$
the expression \re{4.2} for $\chi$ contains the derivative
$$
\frac{d\geff}{d\alpha}=\frac{d\geff}{dc}\left(\frac{d\alpha}{dc}\right)^{-1}
$$
which tends to infinity for the solutions \re{5.6} and \re{5.7} of the equation
$\frac{d\alpha}{dc}=0$ provided that $\frac{d\geff}{dc}\neq 0.$ An
important property of the model is that the derivative of the effective
touching coupling constant $\frac{d\geff}{d\alpha}$ is finite for the
solution \re{5.7} and infinite for \re{5.6}. Indeed, using the expression
\re{3.4} for $\geff$ we obtain
$$
\frac{d\geff}{d\alpha}=\frac{4}{\alpha^2c}
                      +2\left(f^\prime(c)+\frac{2}{\alpha c^2}\right)
                       \left(\frac{d\alpha}{dc}\right)^{-1}
$$
and after substitution of the expression \re{5.5} for $\frac{d\alpha}{dc}$
\be
\frac{d\geff}{d\alpha}=-\frac{4g\alpha c^2\phi(c)}{16-g(\alpha c)^2}
\lab{5.8}
\ee
This relation enables us to identify the peculiarities of the derivative
$\frac{d\geff}{d\alpha}.$ Firstly, equation \re{5.8} is singular for
\re{5.6} and regular for \re{5.7}. Secondly, $\frac{d\geff}{d\alpha}$
depends on the function $\phi(c)$ and it would have additional singularities
if this function is singular for some $c$ as it occurs for $c=-\eta^2$ in
the Penner critical point with the density \re{5.2}.

We forget for a moment about a second possibility and return to its
discussion at the end of the paper. In that case, the singularities of the
derivative $\frac{d\geff}{d\alpha}$ and zeros of $\frac{d\alpha}{dc}$
become correlated and there are only two branches of criticality: \re{5.6}
 and \re{5.7}.
For the solutions \re{5.7} we have $\frac{d\alpha}{dc}=0$ and
$\frac{d\geff}{d\alpha}=\ $finite. It means that the string susceptibility
\re{4.2} is
finite for $\alpha\to\alpha_{cr}$ and the critical exponent is negative
$\gamma_{str} <0.$ For the solutions \re{5.6} the relations
$\frac{d\alpha}{dc}=0,$
$\frac{d\geff}{d\alpha}\to\infty$ lead to the infinite string susceptibility
as $\alpha\to\alpha_{cr}.$ The critical exponent becomes positive and we expect
to get the value $\gamma_{str}=1/2,$ corresponding to the phase of branched
polymers. It turned out that these two branches may cross each other given
rise to intermediate phase with positive critical exponent. To conclude
this section we stress that it is a nonanalyticity in the dependence of
effective touching coupling constant on the cosmological constant which
leads to the appearance of phases with positive critical exponent
$\gamma_{str}.$

\sect{The properties of critical points}

In this section we consider the properties of the solutions \re{5.6} and
\re{5.7}. The equation \re{5.7} relates the critical values
$\alpha=\alpha_{cr}$ and $c=c_{cr}$ for which $\frac{d\alpha}{dc}=0$ and
$$
\alpha-\alpha_{cr} \sim (c-c_{cr})^2.
$$
To value of $c_{cr}$ one gets after substitution of \re{5.7} into
\re{4.3} as a solution of the following equation
\be
4(c_{cr}f^\prime(c_{cr}))^2(c_{cr}f(c_{cr}))^\prime
=g(c_{cr}\phi(c_{cr}))^\prime.
\lab{6.1}
\ee
The effective touching coupling constant at the critical point is found
analogously from \re{3.8}, \re{5.7} and \re{3.6} as
\be
\geff(\alpha_{cr})=2(c_{cr}f(c_{cr}))^\prime
\lab{6.2}
\ee
and its derivative \re{5.8} is given by
\be
\frac{d\geff}{d\alpha}(\alpha_{cr})
=\frac{2gc_{cr}^2\phi(c_{cr})f^\prime(c_{cr})}{4(c_{cr}f^\prime(c_{cr}))^2-g}
\lab{6.3}
\ee
These expressions are finite for all values of $c_{cr}$ except of the
point where $(c_{cr}f^\prime(c_{cr}))^2=\fac14g$ and the equations \re{5.6}
and \re{5.7} coincide. With all these expressions taking into account we
obtain from \re{4.2} the string susceptibility as
$\chi\sim c-c_{cr} \sim (\alpha-\alpha_{cr})^{1/2}.$
In general, by tunning of the density $\rho(x),$ it is possible to satisfy
$\frac{d\alpha}{dc}\sim (c-c_{cr})^{K-1}$ in order to reach the multicritical
points with  $\chi\sim(\alpha-\alpha_{cr})^{1/K}$ and negative exponent
$\gamma_{str}=-1/K.$

Considering the second branch \re{5.6} we note that it exists only for the
positive touching coupling constants $g >0.$ Since the equation \re{5.6} does
not depend on the density $\rho(x)$ it is impossible in contrast with the
previous case to set  higher derivatives $\frac{d^n\alpha}{dc^n},$
$n>1$ to zero  at the critical point and the only possibility remains
$$
\alpha-\alpha_{cr} \sim (c-c_{cr})^2
$$
The critical values $c_{cr}$ and $\alpha_{cr}$ are found from \re{5.6} and
\re{4.3} as solutions of the system
\be
\frac4{\alpha_{cr} c_{cr}}=f(c_{cr})+\phi(c_{cr})=-\sqrt{g}
\lab{6.4}
\ee
The effective touching coupling constant \re{3.8} has a form
\be
\geff(\alpha_{cr})=f(c_{cr})-\phi(c_{cr})=2(c_{cr}\psi(c_{cr}))^\prime
\lab{6.4.1}
\ee
and is finite at the critical point, whereas its derivative \re{5.8} has a
pole as $c\to c_{cr}$
\be
\frac{d\geff(\alpha)}{d\alpha}=\frac{c_{cr}^2\phi(c_{cr})\psi(c_{cr})}
                               {c-c_{cr}}
\lab{6.5}
\ee
Hence, being substituted into \re{4.2}, this derivative defines the asymptotics
of the string susceptibility near the critical point as
\be
\chi\sim-\frac{c_{cr}\phi(c_{cr})}{4\psi(c_{cr})}\frac{d\geff}{d\alpha}
    \sim-\frac{c_{cr}^3\phi^2(c_{cr})}4\frac1{c-c_{cr}}
    \sim (\alpha-\alpha_{cr})^{-1/2}.
\lab{6.5.1}
\ee
The critical exponent $\gamma_{str}=1/2$ corresponds to the phase of branched
polymers.

The solutions of the equations \re{6.1} and \re{6.4} define two curves of
the possible critical behaviour in the $(\alpha,\ g)$ plane. As was shown in
\ci{Touching} in the simplest case of potential $V_0(M),$ these curves are
intersected at some point leading to appearance of new intermediate phase.
The condition for the existence of such point can be found by combining
equations \re{5.7} and \re{6.4} into the system
\be
f(c_{cr})+\phi(c_{cr})=-\sqrt{g},               \qqquad
c_{cr}f^\prime(c_{cr})=\fac12\sqrt{g},          \qqquad
\alpha_{cr}=-\frac4{\sqrt{g}c_{cr}}
\lab{6.6}
\ee
The critical point defined by these equations belongs to both branches of
solutions and has the properties of both. In particular, relations \re{6.3}
and \re{6.5} coincide for the solution of \re{6.6} indicating that the
derivative
$\frac{d\geff}{d\alpha}$ has a pole at the critical point and, as a
consequence, the string susceptibility has asymptotics
$\chi\sim\frac1{c-c_{cr}}.$ Moreover, the both factors in the expression
\re{5.5} for $\frac{d\alpha}{dc}$ equal to zero for \re{6.6}. As only
$\frac{d\alpha}{dc}\sim c-c_{cr}$ on the branch \re{5.6} and
$\frac{d\alpha}{dc}\sim(c-c_{cr})^{K-1}$ after tunning of
parameters of model on the branch \re{5.7}, then at the intersection point one
gets $\frac{d\alpha}{dc}\sim(c-c_{cr})^K$ or
$\alpha-\alpha_{cr}\sim(c-c_{cr})^{K+1}.$ Hence, the string susceptibility
has a behaviour $\chi\sim (\alpha-\alpha_{cr})^{-1/(K+1)}$ with
positive exponent $\gamma_{str}=1/(K+1).$

We conclude that the existence of solutions of the system \re{6.6} would
mean that between the phase of multicritical points with $\gamma_{str}=-1/K$
and
phase of branched polymers with $\gamma_{str}=1/2$ there appears an
intermediate phase of ``hybrid'' multicritical points with the exponent
$\gamma_{str}=1/(K+1).$

\sect{``Hybrid'' multicritical points}

The necessary condition to get the hybrid multicritical point is to satisfy
the relation $\alpha-\alpha_{cr}\sim(c-c_{cr})^K$ near the critical
point defined by equations \re{5.7} and \re{6.1}, or equivalently,
\be
\left.\frac{d\alpha}{dc}\right|_{c=c_{cr}}
=\left.\frac{d^2\alpha}{dc^2}\right|_{c=c_{cr}}
= \cdots
=\left.\frac{d^{K-1}\alpha}{dc^{K-1}}\right|_{c=c_{cr}}
=0
\lab{7.1}
\ee
At the vicinity of the critical point \re{5.7} the derivative
\re{5.5}
is given by
$$
\frac{d\alpha}{dc} \sim f^\prime(c)+\frac2{\alpha c^2}
$$
and condition \re{7.1} can be satisfied under an appropriate choice of function
$f(c).$ Namely, differentiating the both sides of this relation with
respect to $c$ and using \re{7.1}
$$
\left.\frac{d^n\alpha}{dc^n}\right|_{c=c_{cr}}
\sim f^{(n)}(c_{cr})+\frac{2(-1)^{n-1}n!}{\alpha_{cr}c_{cr}^{n+1}}= 0,
\qquad (n=1,\ldots,K-1)
$$
we obtain set equations for the function $f(c)$
$$
f^{(n)}(c_{cr})=\frac{2(-1)^{n}n!}{\alpha_{cr} c_{cr}^{n+1}},
\qquad (n=1,\ldots,K-1)
$$
The remaining higher derivatives are not fixed and the function $f(c)$ is
defined up to $\CO((c-c_{cr})^{K})$ terms. The general form of this
function is
\be
f(c)=A+\frac{2}{\alpha_0 c}\left(1-\left(1-\frac{c}{c_0}\right)^{K}\right)
      +\CO((c-c_0)^{K})
\lab{7.2}
\ee
where arbitrary constants $A$ and $\alpha_0$ are defined as
\be
A=f(c_0)+c_0 f^\prime(c_0)      \qqqquad
\alpha_0=-\frac2{c_0^2f^\prime(c_0)}
\lab{7.2.1}
\ee
and the critical values of $c$ and $\alpha$ are given by
\be
c_{cr}(g)=c_0,                  \qqqquad
\alpha_{cr}(g)=\alpha_0
\lab{7.2.2}
\ee
The expression \re{7.2} has a well defined expansion in powers of $c$ with the
coefficients related to the moments of the density $\rho(x).$ Using
identities \re{3.7} one can evaluate the functions $\phi(c)$ and $\psi(c),$
substitute them into equation \re{6.1} to find
$$
g=4(K+1)\frac{A}{\alpha_0c_0}
$$
This equation implies that the constants $A,$ $\alpha_0$ and $c_0,$ as well as
the critical values of $c_{cr}$ and $\alpha_{cr}$ are functions of
touching coupling constant $g.$ After their substitution into
\re{7.2} the function $f(c)$ also becomes $g$ dependent. However,
according to the definition \re{2.7} of the model the density $\rho(x)$ and
function $f(c)$ do not depend on $g.$ The identity
$
\frac{df(c)}{dg} = 0
$
leads to additional restrictions on the possible form of $f(c).$
After substitution of \re{7.2} this equation has the following two solutions
\be
\frac{d c_0(g)}{d g}=\frac{d \alpha_0(g)}{d g}=\frac{d A(g)}{d g}=0
\lab{7.3}
\ee
for an arbitrary $K\geq 2$ and the second solution exists only for the special
value of $K$
\be
\frac{d A(g)}{d c_0}=-\frac{4}{\alpha_0(g)c_0^2(g)},       \qqqquad
\frac{d \alpha_0(g)}{d c_0}=-\frac{2\alpha_0(g)}{c_0(g)},  \qquad
\mbox{for $K=2$}
\lab{7.4}
\ee
For the last solution $f(c)$ is a linear function of $c$ and the
corresponding potential has a form $V_0(M)\sim M^2 + a M^4$ with some $a.$
It is this model that was considered in \ci{Touching}. Notice, that
in order to get multicritical points with $K\geq 2$ one has to consider
only first solution \re{7.3}. Relation \re{7.3} implies
that on the branch \re{5.7} of criticality the critical values $c_{cr}$ and
$\alpha_{cr}$ do not depend on the touching coupling constant $g.$ On the
other hand, the value of $c_{cr}$ one gets solving equation \re{6.1} which
depends on $g.$ The only way to satisfy the both conditions is to set
$$
\left.(cf(c))^\prime\right|_{c=c_{0}}=c_0f^\prime(c_0)+f(c_0)=0
$$
or to put $A=0$ in expression \re{7.2} for $f(c).$ The resulting function
$f(c)$ coincides (for $\alpha_0=1$ and $c_0=-4$) with an analogous function
\re{5.4} leading to Kazakov's multicritical points for $g=0.$ However, for
$g\neq 0$ equation \re{6.1} leads to an additional constraint on the density
$\rho(x)$
$$
\left.(c\phi(c))^\prime\right|_{c=c_{0}}=c_0\phi^\prime(c_0)+\phi(c_0)=0
$$
or using \re{3.6}
\be
f(c_0)=\phi(c_0) \qquad f^\prime(c_0)=\phi^\prime(c_0)
\lab{7.5}
\ee
It can be easily verified that these conditions are not fulfilled if one
chooses for $f(c)$ the expression \re{7.2} (with $A=0$) without
$\CO((c-c_0)^{K})$ terms. The ``minimal'' expression consistent with the
condition \re{7.5} one gets by adding to $f(c)$ the following term:
$\frac{(K+2)B}{c_0}\left(1-\frac{c}{c_0}\right)^{K}$ and after
simple calculations
\ba
cf(c) &=& \frac2{\alpha_0}\left( 1+(K+1)\left(1-\frac{c}{c_0}\right)^{K}
                    -(K+2)\left(1-\frac{c}{c_0}\right)^{K+1}
            \right)
\nonumber \\
c\phi(c) &=& \frac2{\alpha_0}\left( 1-(K+1)\left(1-\frac{c}{c_0}\right)^{K}
                    +K\left(1-\frac{c}{c_0}\right)^{K+1}
            \right)
\lab{7.6} \\
c\psi(c) &=& \frac2{\alpha_0}\left( 1-\left(1-\frac{c}{c_0}\right)^{K+1}
            \right)
\nonumber
\ea
Using the definition \re{3.5} of the function $f(c)$ we calculate the
moments of the density and then find the potential $V_0(M)$ as
\be
V_0(M)=-\frac1{\alpha_0}\left(
                            (K+1)U_{2K}(M^\prime)-(K+2)U_{2(K+1)}(M^\prime)
                        \right)
\lab{7.7}
\ee
where $M^\prime\equiv\frac{2}{\sqrt{-c_0}}M$ and $U_{2K}(M)$ is the potential
\re{2.3} of the $K-$th multicritical Kazakov's model. It is this potential
which leads to the appearance of the hybrid multicritical points.

Let us consider the critical behaviour of the model with the potential
\re{7.7} following the general analysis of the previous section.

\bigskip

\noindent
{\it { Smooth surfaces:}} $g < g_0$

\bigskip

\noindent
For $g=0$ the potential \re{2.7} is given by \re{7.7} and the model has
a critical behaviour corresponding to the  $K-$th multicritical point. The
values of the critical parameters are given by \re{7.2.2}
and they lie on the branch \re{5.7} of solutions. Moreover, for $g < 0$ this
branch is the unique. After increasing of $g > 0$ the model follows the
same branch as for $g=0$ until the touching constant reaches the critical
value $g_0$ at which the both solutions \re{5.6} and \re{5.7} coincide.
The value $g_0$ is found from the system \re{6.6} and \re{7.6} as
$$
g_0 = \frac{16}{(\alpha_0c_0)^2}
$$
Thus, for $g < g_0$ the critical behaviour of the model is described by
the equations \re{5.7}, \re{6.1} and \re{7.6}. This phase has the following
remarkable properties. It follows from \re{7.2.2} and \re{7.3} that the
critical
values $\alpha_{cr}$ and $c_{cr}$ don't depend on $g$ and coincide with
analogous values \re{7.2.2} for the $K-$th multicritical model without
touchings. Substituting \re{7.2.2} and \re{7.6} into \re{6.2} we find that
the effective touching coupling constant vanishes at the critical point
\be
\geff(\alpha_{cr})=0, \qquad \mbox{for $g < g_0$}
\lab{7.9}
\ee
To get the behaviour of $\alpha$ and $\geff$ near the critical point
\re{7.2.2} one evaluates their derivatives \re{5.5} and \re{6.3} and
gets
$$
\left(1-\frac{\alpha}{\alpha_0}\right)=(K+1)\frac{g_0-g}{g_0+g}
\left(1-\frac{c}{c_0}\right)^{K}+\CO\left((c-c_0)^{K+1}\right)
$$
and
\be
\geff(\alpha)=-\frac{c_0}2\frac{g_0g}{g_0-g}(\alpha-\alpha_0)
+\CO\left((\alpha-\alpha_0)^{\frac{K+1}{K}}\right)
\lab{7.10}
\ee
Note, that $\geff$ is not analytical in $\alpha$ near the critical point
$\alpha_{cr}=\alpha_0$ and analytical in $g$ for $g < g_0.$ It means that
at the vicinity of the critical point the effective touching constant
$\geff$ is defined by the random surfaces with infinite area and only
finite number of touchings. One uses \re{4.2} and \re{7.10}
to calculate the derivative
$\left.\frac{d\chi}{dc}\right|_{c=c_{cr}}=\frac{1}{\alpha_0c_0}
 \left(\frac{g_0+g}{g_0-g}\right)^2\neq 0.$ Hence,
the string susceptibility \re{4.2} in this phase is given by
$$
\chi\sim (c-c_0)\sim (\alpha-\alpha_0)^{1/K}
$$
indicating that for $g < g_0$ the model looks like ordinary $K-$th
multicritical model.

\bigskip

\noindent
 {\it { Intermediate phase:}} $g = g_0.$

\bigskip

\noindent
For this value of $g$ the solutions \re{5.6} and \re{5.7} coincide giving
rise to new critical point with
\be
c_{cr}(g_0)=c_0,                        \qqquad
\alpha_{cr}(g_0)=\alpha_0               \qqquad
g=g_0=\frac{16}{(\alpha_0c_0)^2}
\lab{7.11}
\ee
To find the scaling of parameters near the critical point one
substitutes \re{7.6} and \re{7.11} into \re{5.5} and obtains
$$
\frac{d\alpha}{dc}=K(K+1)\frac{\alpha_0}{c_0}
\left(1-\frac{c}{c_0}\right)^{K}+\CO\left((c-c_0)^{K+1}\right)
$$
and integrating
$$
\left(1-\frac{\alpha}{\alpha_0}\right)
=K\left(1-\frac{c}{c_0}\right)^{K+1}+\CO\left((c-c_0)^{K+2}\right)
$$
The effective touching constant is found analogously from \re{3.8} as
\be
\geff(\alpha)=\frac{4}{\alpha_0c_0}(K+1)K^{-\frac{K}{K+1}}
\left(1-\frac{\alpha}{\alpha_0}\right)^{\frac{K}{K+1}}
\lab{7.12}
\ee
Although it vanishes at the critical point \re{7.11}, the derivative
$\frac{d\geff}{d\alpha}\to\infty$ tends to infinity.
The asymptotics of the string susceptibility \re{4.2} in this phase is
determined by the singularity of the derivative $\frac{d\geff}{d\alpha}$
$$
\chi = -\frac18\alpha_0c_0^2\psi(c_0)\frac{d\geff}{d\alpha}
     \sim (\alpha-\alpha_0)^{-1/(K+1)}
$$
with positive critical exponent $\gamma_{str}=1/(K+1).$

\bigskip

\noindent
{\it { Branched polymers:}} $g_0 < g < g_{max}$

\bigskip

\noindent
For large values of the touching constant the model passes to the phase of
branched polymers described by the relations \re{6.4} and \re{7.6}. The
equation \re{6.4} for critical point looks like
$$
\left(1-\frac{c_{cr}(g)}{c_0}\right)^{K+1}=1-\frac{\alpha_0}{\alpha_{cr}(g)}
$$
and after it substitution into \re{6.4} and \re{6.4.1}
\be
\left(\frac{g_0}{g}\right)^{\frac12}=\frac{\alpha_{cr}(g)}{\alpha_0}
\left(1-\left(1-\frac{\alpha_0}{\alpha_{cr}(g)}\right)^{\frac1{K+1}}\right)
\lab{7.13}
\ee
and
$$
\geff(\alpha_{cr})=\frac{4(K+1)}{\alpha_0c_0}
\left(1-\frac{\alpha_0}{\alpha_{cr}(g)}\right)^\frac{K}{K+1}
$$
These relations give us the dependence of the critical values of
cosmological constant $\alpha_{cr}$ and effective touching coupling
constant $\geff(\alpha_{cr})$ on $g$ as shown in fig.~1.
We notice that for $g > g_0$ the critical point starts to depend on the
touching constant. In the phase of branched polymers the critical value
$\alpha_{cr}$ is increasing function of touching constant and in the limit
when $g$ approaches a maximum value the cosmological constant tends to
infinity
$$
g \to g_{max} = g_0 (K+1)^2,     \qqqquad
\alpha_{cr}(g)\to\infty,            \qqqquad
\geff(\alpha_{cr})\to\frac{4(K+1)}{\alpha_0c_0}
$$
In an opposite limit, $g \to g_0,$ the dependence of $\alpha_{cr}$
and $\geff(\alpha_{cr})$ on $g$ is found from \re{7.13} as
$$
\left(\frac{\alpha_{cr}(g)}{\alpha_0}-1\right)
=2^{-K-1}\left(\frac{g}{g_0}-1\right)^{K+1},
\qqquad
\geff(\alpha_{cr})=\frac{K+1}{2^{K-2}\alpha_0c_0}\left(\frac{g}{g_0}-1\right)^K
$$
These relations imply that under transition of the system through the
critical point with $g=g_0$ and $\alpha_{cr}=\alpha_0$ the derivatives
$\frac{d^{K}\alpha}{dg^{K}}$ and $\frac{d^{K-1}\geff}{dg^{K-1}}$
undergo a jump.

Let us consider the behaviour of parameters near the line
$\alpha_{cr}=\alpha_{cr}(g)$ of criticality. Since
$\left.\frac{d\alpha}{dc}\right|_{\alpha=\alpha_{cr}}=0$ and
$\left.\frac{d^2\alpha}{dc^2}\right|_{\alpha=\alpha_{cr}}\neq 0$
then
$
\alpha-\alpha_{cr}\sim(c-c_{cr})^2.
$
As follows from \re{6.5},
$
\frac{d\geff}{d\alpha}=-\frac{\sqrt{g}}2\frac{c_{cr}\phi(c_{cr})}{c-c_{cr}}
\sim \frac1{\sqrt{\alpha-\alpha_{cr}}}
$
and the derivative of $\geff$ tends to infinity as $\alpha\to\alpha_{cr}.$
Integration of this relation gives
\be
\geff(\alpha) - \geff(\alpha_{cr}) \sim \sqrt{\alpha-\alpha_{cr}}.
\lab{7.14}
\ee
As was mentioned before, the string susceptibility in this phase is equal
to \re{6.5.1} with the exponent $\gamma_{str}=1/2$ corresponding to the
branched
polymers.

\hspace{-20mm}
\noindent
\setlength{\unitlength}{0.240900pt}
\begin{picture}(1049,629)(0,0)
\tenrm
\ifx\plotpoint\undefined\newsavebox{\plotpoint}\fi
\put(242,113){\makebox(0,0)[r]{0}}
\put(206,561){\makebox(0,0)[l]{$g$}}
\put(206,256){\makebox(0,0)[l]{$g_0$}}
\put(956,100){\makebox(0,0)[l]{$\alpha$}}
\put(437,100){\makebox(0,0)[l]{$\alpha_0$}}
\sbox{\plotpoint}{\rule[-0.200pt]{1.400pt}{1.400pt}}%
\put(206,133){\vector(1,0){808}}
\put(264,93){\vector(0,1){488}}
\put(437,256){\usebox{\plotpoint}}
\put(437,256){\rule[-0.200pt]{1.400pt}{4.156pt}}
\put(438,273){\rule[-0.200pt]{1.400pt}{4.156pt}}
\put(439,290){\rule[-0.200pt]{1.400pt}{4.156pt}}
\put(440,307){\rule[-0.200pt]{1.400pt}{4.156pt}}
\put(441,325){\rule[-0.200pt]{1.400pt}{2.329pt}}
\put(442,334){\rule[-0.200pt]{1.400pt}{2.329pt}}
\put(443,344){\rule[-0.200pt]{1.400pt}{2.329pt}}
\put(444,353){\rule[-0.200pt]{1.400pt}{1.265pt}}
\put(445,359){\rule[-0.200pt]{1.400pt}{1.265pt}}
\put(446,364){\rule[-0.200pt]{1.400pt}{1.265pt}}
\put(447,369){\rule[-0.200pt]{1.400pt}{1.265pt}}
\put(448,375){\rule[-0.200pt]{1.400pt}{1.365pt}}
\put(449,380){\rule[-0.200pt]{1.400pt}{1.365pt}}
\put(450,386){\rule[-0.200pt]{1.400pt}{1.365pt}}
\put(451,391){\rule[-0.200pt]{1.400pt}{0.843pt}}
\put(452,395){\rule[-0.200pt]{1.400pt}{0.843pt}}
\put(453,399){\rule[-0.200pt]{1.400pt}{0.843pt}}
\put(454,402){\rule[-0.200pt]{1.400pt}{0.843pt}}
\put(455,406){\rule[-0.200pt]{1.400pt}{0.723pt}}
\put(456,409){\rule[-0.200pt]{1.400pt}{0.723pt}}
\put(457,412){\rule[-0.200pt]{1.400pt}{0.723pt}}
\put(458,415){\rule[-0.200pt]{1.400pt}{0.723pt}}
\put(459,418){\rule[-0.200pt]{1.400pt}{0.803pt}}
\put(460,421){\rule[-0.200pt]{1.400pt}{0.803pt}}
\put(461,424){\rule[-0.200pt]{1.400pt}{0.803pt}}
\put(462,428){\rule[-0.200pt]{1.400pt}{0.482pt}}
\put(463,430){\rule[-0.200pt]{1.400pt}{0.482pt}}
\put(464,432){\rule[-0.200pt]{1.400pt}{0.482pt}}
\put(465,434){\rule[-0.200pt]{1.400pt}{0.482pt}}
\put(466,436){\rule[-0.200pt]{1.400pt}{0.642pt}}
\put(467,438){\rule[-0.200pt]{1.400pt}{0.642pt}}
\put(468,441){\rule[-0.200pt]{1.400pt}{0.642pt}}
\put(469,443){\rule[-0.200pt]{1.400pt}{0.422pt}}
\put(470,445){\rule[-0.200pt]{1.400pt}{0.422pt}}
\put(471,447){\rule[-0.200pt]{1.400pt}{0.422pt}}
\put(472,449){\rule[-0.200pt]{1.400pt}{0.422pt}}
\put(473,451){\usebox{\plotpoint}}
\put(474,452){\usebox{\plotpoint}}
\put(475,454){\usebox{\plotpoint}}
\put(476,455){\usebox{\plotpoint}}
\put(477,457){\rule[-0.200pt]{1.400pt}{0.401pt}}
\put(478,458){\rule[-0.200pt]{1.400pt}{0.401pt}}
\put(479,460){\rule[-0.200pt]{1.400pt}{0.401pt}}
\put(480,461){\usebox{\plotpoint}}
\put(481,463){\usebox{\plotpoint}}
\put(482,464){\usebox{\plotpoint}}
\put(483,465){\usebox{\plotpoint}}
\put(484,467){\usebox{\plotpoint}}
\put(485,468){\usebox{\plotpoint}}
\put(486,469){\usebox{\plotpoint}}
\put(487,470){\usebox{\plotpoint}}
\put(488,471){\usebox{\plotpoint}}
\put(489,472){\usebox{\plotpoint}}
\put(490,473){\usebox{\plotpoint}}
\put(491,475){\usebox{\plotpoint}}
\put(492,476){\usebox{\plotpoint}}
\put(493,477){\usebox{\plotpoint}}
\put(494,478){\usebox{\plotpoint}}
\put(495,479){\usebox{\plotpoint}}
\put(496,480){\usebox{\plotpoint}}
\put(497,481){\usebox{\plotpoint}}
\put(498,482){\usebox{\plotpoint}}
\put(499,483){\usebox{\plotpoint}}
\put(500,484){\usebox{\plotpoint}}
\put(502,485){\usebox{\plotpoint}}
\put(503,486){\usebox{\plotpoint}}
\put(504,487){\usebox{\plotpoint}}
\put(506,488){\usebox{\plotpoint}}
\put(507,489){\usebox{\plotpoint}}
\put(508,490){\usebox{\plotpoint}}
\put(509,491){\rule[-0.200pt]{0.482pt}{1.400pt}}
\put(511,492){\rule[-0.200pt]{0.482pt}{1.400pt}}
\put(513,493){\usebox{\plotpoint}}
\put(514,494){\usebox{\plotpoint}}
\put(516,495){\rule[-0.200pt]{0.482pt}{1.400pt}}
\put(518,496){\rule[-0.200pt]{0.482pt}{1.400pt}}
\put(520,497){\rule[-0.200pt]{0.482pt}{1.400pt}}
\put(522,498){\rule[-0.200pt]{0.482pt}{1.400pt}}
\put(524,499){\usebox{\plotpoint}}
\put(525,500){\usebox{\plotpoint}}
\put(527,501){\rule[-0.200pt]{0.482pt}{1.400pt}}
\put(529,502){\rule[-0.200pt]{0.482pt}{1.400pt}}
\put(531,503){\usebox{\plotpoint}}
\put(532,504){\usebox{\plotpoint}}
\put(534,505){\rule[-0.200pt]{0.964pt}{1.400pt}}
\put(538,506){\rule[-0.200pt]{0.482pt}{1.400pt}}
\put(540,507){\rule[-0.200pt]{0.482pt}{1.400pt}}
\put(542,508){\rule[-0.200pt]{0.723pt}{1.400pt}}
\put(545,509){\rule[-0.200pt]{0.482pt}{1.400pt}}
\put(547,510){\rule[-0.200pt]{0.482pt}{1.400pt}}
\put(549,511){\rule[-0.200pt]{0.723pt}{1.400pt}}
\put(552,512){\rule[-0.200pt]{0.964pt}{1.400pt}}
\put(556,513){\rule[-0.200pt]{0.964pt}{1.400pt}}
\put(560,514){\rule[-0.200pt]{0.723pt}{1.400pt}}
\put(563,515){\rule[-0.200pt]{0.964pt}{1.400pt}}
\put(567,516){\rule[-0.200pt]{0.723pt}{1.400pt}}
\put(570,517){\rule[-0.200pt]{0.964pt}{1.400pt}}
\put(574,518){\rule[-0.200pt]{0.964pt}{1.400pt}}
\put(578,519){\rule[-0.200pt]{0.723pt}{1.400pt}}
\put(581,520){\rule[-0.200pt]{0.964pt}{1.400pt}}
\put(585,521){\rule[-0.200pt]{0.723pt}{1.400pt}}
\put(588,522){\rule[-0.200pt]{0.964pt}{1.400pt}}
\put(592,523){\rule[-0.200pt]{1.686pt}{1.400pt}}
\put(599,524){\rule[-0.200pt]{0.964pt}{1.400pt}}
\put(603,525){\rule[-0.200pt]{1.686pt}{1.400pt}}
\put(610,526){\rule[-0.200pt]{0.964pt}{1.400pt}}
\put(614,527){\rule[-0.200pt]{1.686pt}{1.400pt}}
\put(621,528){\rule[-0.200pt]{1.686pt}{1.400pt}}
\put(628,529){\rule[-0.200pt]{1.686pt}{1.400pt}}
\put(635,530){\rule[-0.200pt]{1.927pt}{1.400pt}}
\put(643,531){\rule[-0.200pt]{1.686pt}{1.400pt}}
\put(650,532){\rule[-0.200pt]{1.686pt}{1.400pt}}
\put(657,533){\rule[-0.200pt]{2.650pt}{1.400pt}}
\put(668,534){\rule[-0.200pt]{1.686pt}{1.400pt}}
\put(675,535){\rule[-0.200pt]{2.650pt}{1.400pt}}
\put(686,536){\rule[-0.200pt]{2.650pt}{1.400pt}}
\put(697,537){\rule[-0.200pt]{3.373pt}{1.400pt}}
\put(711,538){\rule[-0.200pt]{3.373pt}{1.400pt}}
\put(725,539){\rule[-0.200pt]{3.614pt}{1.400pt}}
\put(740,540){\rule[-0.200pt]{3.373pt}{1.400pt}}
\put(754,541){\rule[-0.200pt]{4.336pt}{1.400pt}}
\put(772,542){\rule[-0.200pt]{4.336pt}{1.400pt}}
\put(790,543){\rule[-0.200pt]{5.300pt}{1.400pt}}
\put(812,544){\rule[-0.200pt]{6.022pt}{1.400pt}}
\put(837,545){\rule[-0.200pt]{6.986pt}{1.400pt}}
\put(866,546){\rule[-0.200pt]{7.709pt}{1.400pt}}
\put(898,547){\rule[-0.200pt]{7.950pt}{1.400pt}}
\put(931,548){\rule[-0.200pt]{10.359pt}{1.400pt}}
\put(974,549){\rule[-0.200pt]{2.650pt}{1.400pt}}
\sbox{\plotpoint}{\rule[-0.500pt]{1.400pt}{1.400pt}}%
\put(437,133){\usebox{\plotpoint}}
\put(437,133){\rule[-0.500pt]{1.400pt}{29.631pt}}
\sbox{\plotpoint}{\rule[-1.000pt]{2.000pt}{2.000pt}}%
\put(437,256){\circle*{30}}
\end{picture}
\setlength{\unitlength}{0.240900pt}
\begin{picture}(1049,629)(0,0)
\tenrm
\ifx\plotpoint\undefined\newsavebox{\plotpoint}\fi
\put(242,113){\makebox(0,0)[r]{0}}
\put(985,100){\makebox(0,0){$\geff$}}
\put(192,561){\makebox(0,0)[l]{$g$}}
\put(192,256){\makebox(0,0)[l]{$g_0$}}
\sbox{\plotpoint}{\rule[-0.200pt]{1.400pt}{1.400pt}}%
\put(192,133){\vector(1,0){825}}
\put(264,93){\vector(0,1){488}}
\put(264,256){\usebox{\plotpoint}}
\put(264,256){\rule[-0.200pt]{1.400pt}{0.482pt}}
\put(265,258){\rule[-0.200pt]{1.400pt}{0.482pt}}
\put(266,260){\rule[-0.200pt]{1.400pt}{0.482pt}}
\put(267,262){\rule[-0.200pt]{1.400pt}{0.482pt}}
\put(268,264){\rule[-0.200pt]{1.400pt}{0.482pt}}
\put(269,266){\usebox{\plotpoint}}
\put(270,267){\usebox{\plotpoint}}
\put(271,268){\usebox{\plotpoint}}
\put(272,269){\usebox{\plotpoint}}
\put(273,270){\rule[-0.200pt]{0.401pt}{1.400pt}}
\put(274,271){\rule[-0.200pt]{0.401pt}{1.400pt}}
\put(276,272){\rule[-0.200pt]{0.401pt}{1.400pt}}
\put(277,273){\rule[-0.200pt]{0.482pt}{1.400pt}}
\put(280,274){\rule[-0.200pt]{0.482pt}{1.400pt}}
\put(282,275){\rule[-0.200pt]{0.401pt}{1.400pt}}
\put(283,276){\rule[-0.200pt]{0.401pt}{1.400pt}}
\put(285,277){\rule[-0.200pt]{0.401pt}{1.400pt}}
\put(286,278){\rule[-0.200pt]{0.482pt}{1.400pt}}
\put(289,279){\rule[-0.200pt]{0.482pt}{1.400pt}}
\put(291,280){\rule[-0.200pt]{0.602pt}{1.400pt}}
\put(293,281){\rule[-0.200pt]{0.602pt}{1.400pt}}
\put(296,282){\rule[-0.200pt]{0.482pt}{1.400pt}}
\put(298,283){\rule[-0.200pt]{0.482pt}{1.400pt}}
\put(300,284){\rule[-0.200pt]{0.602pt}{1.400pt}}
\put(302,285){\rule[-0.200pt]{0.602pt}{1.400pt}}
\put(305,286){\rule[-0.200pt]{0.482pt}{1.400pt}}
\put(307,287){\rule[-0.200pt]{0.482pt}{1.400pt}}
\put(309,288){\rule[-0.200pt]{0.602pt}{1.400pt}}
\put(311,289){\rule[-0.200pt]{0.602pt}{1.400pt}}
\put(314,290){\rule[-0.200pt]{0.482pt}{1.400pt}}
\put(316,291){\rule[-0.200pt]{0.482pt}{1.400pt}}
\put(318,292){\rule[-0.200pt]{0.602pt}{1.400pt}}
\put(320,293){\rule[-0.200pt]{0.602pt}{1.400pt}}
\put(323,294){\rule[-0.200pt]{0.482pt}{1.400pt}}
\put(325,295){\rule[-0.200pt]{0.482pt}{1.400pt}}
\put(327,296){\rule[-0.200pt]{0.602pt}{1.400pt}}
\put(329,297){\rule[-0.200pt]{0.602pt}{1.400pt}}
\put(332,298){\rule[-0.200pt]{0.482pt}{1.400pt}}
\put(334,299){\rule[-0.200pt]{0.482pt}{1.400pt}}
\put(336,300){\rule[-0.200pt]{1.204pt}{1.400pt}}
\put(341,301){\rule[-0.200pt]{0.482pt}{1.400pt}}
\put(343,302){\rule[-0.200pt]{0.482pt}{1.400pt}}
\put(345,303){\rule[-0.200pt]{0.602pt}{1.400pt}}
\put(347,304){\rule[-0.200pt]{0.602pt}{1.400pt}}
\put(350,305){\rule[-0.200pt]{0.482pt}{1.400pt}}
\put(352,306){\rule[-0.200pt]{0.482pt}{1.400pt}}
\put(354,307){\rule[-0.200pt]{0.602pt}{1.400pt}}
\put(356,308){\rule[-0.200pt]{0.602pt}{1.400pt}}
\put(359,309){\rule[-0.200pt]{0.964pt}{1.400pt}}
\put(363,310){\rule[-0.200pt]{0.602pt}{1.400pt}}
\put(365,311){\rule[-0.200pt]{0.602pt}{1.400pt}}
\put(368,312){\rule[-0.200pt]{0.482pt}{1.400pt}}
\put(370,313){\rule[-0.200pt]{0.482pt}{1.400pt}}
\put(372,314){\rule[-0.200pt]{1.204pt}{1.400pt}}
\put(377,315){\rule[-0.200pt]{0.482pt}{1.400pt}}
\put(379,316){\rule[-0.200pt]{0.482pt}{1.400pt}}
\put(381,317){\rule[-0.200pt]{0.602pt}{1.400pt}}
\put(383,318){\rule[-0.200pt]{0.602pt}{1.400pt}}
\put(386,319){\rule[-0.200pt]{0.482pt}{1.400pt}}
\put(388,320){\rule[-0.200pt]{0.482pt}{1.400pt}}
\put(390,321){\rule[-0.200pt]{1.204pt}{1.400pt}}
\put(395,322){\rule[-0.200pt]{0.482pt}{1.400pt}}
\put(397,323){\rule[-0.200pt]{0.482pt}{1.400pt}}
\put(399,324){\rule[-0.200pt]{0.602pt}{1.400pt}}
\put(401,325){\rule[-0.200pt]{0.602pt}{1.400pt}}
\put(404,326){\rule[-0.200pt]{0.964pt}{1.400pt}}
\put(408,327){\rule[-0.200pt]{0.602pt}{1.400pt}}
\put(410,328){\rule[-0.200pt]{0.602pt}{1.400pt}}
\put(413,329){\rule[-0.200pt]{0.482pt}{1.400pt}}
\put(415,330){\rule[-0.200pt]{0.482pt}{1.400pt}}
\put(417,331){\rule[-0.200pt]{1.204pt}{1.400pt}}
\put(422,332){\rule[-0.200pt]{0.482pt}{1.400pt}}
\put(424,333){\rule[-0.200pt]{0.482pt}{1.400pt}}
\put(426,334){\rule[-0.200pt]{0.602pt}{1.400pt}}
\put(428,335){\rule[-0.200pt]{0.602pt}{1.400pt}}
\put(431,336){\rule[-0.200pt]{0.964pt}{1.400pt}}
\put(435,337){\rule[-0.200pt]{0.602pt}{1.400pt}}
\put(437,338){\rule[-0.200pt]{0.602pt}{1.400pt}}
\put(440,339){\rule[-0.200pt]{0.482pt}{1.400pt}}
\put(442,340){\rule[-0.200pt]{0.482pt}{1.400pt}}
\put(444,341){\rule[-0.200pt]{1.204pt}{1.400pt}}
\put(449,342){\rule[-0.200pt]{0.482pt}{1.400pt}}
\put(451,343){\rule[-0.200pt]{0.482pt}{1.400pt}}
\put(453,344){\rule[-0.200pt]{0.602pt}{1.400pt}}
\put(455,345){\rule[-0.200pt]{0.602pt}{1.400pt}}
\put(458,346){\rule[-0.200pt]{0.964pt}{1.400pt}}
\put(462,347){\rule[-0.200pt]{0.602pt}{1.400pt}}
\put(464,348){\rule[-0.200pt]{0.602pt}{1.400pt}}
\put(467,349){\rule[-0.200pt]{0.482pt}{1.400pt}}
\put(469,350){\rule[-0.200pt]{0.482pt}{1.400pt}}
\put(471,351){\rule[-0.200pt]{0.602pt}{1.400pt}}
\put(473,352){\rule[-0.200pt]{0.602pt}{1.400pt}}
\put(476,353){\rule[-0.200pt]{0.964pt}{1.400pt}}
\put(480,354){\rule[-0.200pt]{0.602pt}{1.400pt}}
\put(482,355){\rule[-0.200pt]{0.602pt}{1.400pt}}
\put(485,356){\rule[-0.200pt]{0.482pt}{1.400pt}}
\put(487,357){\rule[-0.200pt]{0.482pt}{1.400pt}}
\put(489,358){\rule[-0.200pt]{1.204pt}{1.400pt}}
\put(494,359){\rule[-0.200pt]{0.482pt}{1.400pt}}
\put(496,360){\rule[-0.200pt]{0.482pt}{1.400pt}}
\put(498,361){\rule[-0.200pt]{0.602pt}{1.400pt}}
\put(500,362){\rule[-0.200pt]{0.602pt}{1.400pt}}
\put(503,363){\rule[-0.200pt]{0.964pt}{1.400pt}}
\put(507,364){\rule[-0.200pt]{0.602pt}{1.400pt}}
\put(509,365){\rule[-0.200pt]{0.602pt}{1.400pt}}
\put(512,366){\rule[-0.200pt]{0.482pt}{1.400pt}}
\put(514,367){\rule[-0.200pt]{0.482pt}{1.400pt}}
\put(516,368){\rule[-0.200pt]{1.204pt}{1.400pt}}
\put(521,369){\rule[-0.200pt]{0.482pt}{1.400pt}}
\put(523,370){\rule[-0.200pt]{0.482pt}{1.400pt}}
\put(525,371){\rule[-0.200pt]{0.602pt}{1.400pt}}
\put(527,372){\rule[-0.200pt]{0.602pt}{1.400pt}}
\put(530,373){\rule[-0.200pt]{0.482pt}{1.400pt}}
\put(532,374){\rule[-0.200pt]{0.482pt}{1.400pt}}
\put(534,375){\rule[-0.200pt]{1.204pt}{1.400pt}}
\put(539,376){\rule[-0.200pt]{0.482pt}{1.400pt}}
\put(541,377){\rule[-0.200pt]{0.482pt}{1.400pt}}
\put(543,378){\rule[-0.200pt]{0.602pt}{1.400pt}}
\put(545,379){\rule[-0.200pt]{0.602pt}{1.400pt}}
\put(548,380){\rule[-0.200pt]{0.964pt}{1.400pt}}
\put(552,381){\rule[-0.200pt]{0.602pt}{1.400pt}}
\put(554,382){\rule[-0.200pt]{0.602pt}{1.400pt}}
\put(557,383){\rule[-0.200pt]{0.482pt}{1.400pt}}
\put(559,384){\rule[-0.200pt]{0.482pt}{1.400pt}}
\put(561,385){\rule[-0.200pt]{0.602pt}{1.400pt}}
\put(563,386){\rule[-0.200pt]{0.602pt}{1.400pt}}
\put(566,387){\rule[-0.200pt]{0.964pt}{1.400pt}}
\put(570,388){\rule[-0.200pt]{0.602pt}{1.400pt}}
\put(572,389){\rule[-0.200pt]{0.602pt}{1.400pt}}
\put(575,390){\rule[-0.200pt]{0.482pt}{1.400pt}}
\put(577,391){\rule[-0.200pt]{0.482pt}{1.400pt}}
\put(579,392){\rule[-0.200pt]{1.204pt}{1.400pt}}
\put(584,393){\rule[-0.200pt]{0.482pt}{1.400pt}}
\put(586,394){\rule[-0.200pt]{0.482pt}{1.400pt}}
\put(588,395){\rule[-0.200pt]{0.602pt}{1.400pt}}
\put(590,396){\rule[-0.200pt]{0.602pt}{1.400pt}}
\put(593,397){\rule[-0.200pt]{0.482pt}{1.400pt}}
\put(595,398){\rule[-0.200pt]{0.482pt}{1.400pt}}
\put(597,399){\rule[-0.200pt]{1.204pt}{1.400pt}}
\put(602,400){\rule[-0.200pt]{0.482pt}{1.400pt}}
\put(604,401){\rule[-0.200pt]{0.482pt}{1.400pt}}
\put(606,402){\rule[-0.200pt]{0.602pt}{1.400pt}}
\put(608,403){\rule[-0.200pt]{0.602pt}{1.400pt}}
\put(611,404){\rule[-0.200pt]{0.482pt}{1.400pt}}
\put(613,405){\rule[-0.200pt]{0.482pt}{1.400pt}}
\put(615,406){\rule[-0.200pt]{1.204pt}{1.400pt}}
\put(620,407){\rule[-0.200pt]{0.602pt}{1.400pt}}
\put(622,408){\rule[-0.200pt]{0.602pt}{1.400pt}}
\put(625,409){\rule[-0.200pt]{0.482pt}{1.400pt}}
\put(627,410){\rule[-0.200pt]{0.482pt}{1.400pt}}
\put(629,411){\rule[-0.200pt]{0.602pt}{1.400pt}}
\put(631,412){\rule[-0.200pt]{0.602pt}{1.400pt}}
\put(634,413){\rule[-0.200pt]{0.482pt}{1.400pt}}
\put(636,414){\rule[-0.200pt]{0.482pt}{1.400pt}}
\put(638,415){\rule[-0.200pt]{1.204pt}{1.400pt}}
\put(643,416){\rule[-0.200pt]{0.482pt}{1.400pt}}
\put(645,417){\rule[-0.200pt]{0.482pt}{1.400pt}}
\put(647,418){\rule[-0.200pt]{0.602pt}{1.400pt}}
\put(649,419){\rule[-0.200pt]{0.602pt}{1.400pt}}
\put(652,420){\rule[-0.200pt]{0.482pt}{1.400pt}}
\put(654,421){\rule[-0.200pt]{0.482pt}{1.400pt}}
\put(656,422){\rule[-0.200pt]{1.204pt}{1.400pt}}
\put(661,423){\rule[-0.200pt]{0.482pt}{1.400pt}}
\put(663,424){\rule[-0.200pt]{0.482pt}{1.400pt}}
\put(665,425){\rule[-0.200pt]{0.602pt}{1.400pt}}
\put(667,426){\rule[-0.200pt]{0.602pt}{1.400pt}}
\put(670,427){\rule[-0.200pt]{0.482pt}{1.400pt}}
\put(672,428){\rule[-0.200pt]{0.482pt}{1.400pt}}
\put(674,429){\rule[-0.200pt]{0.602pt}{1.400pt}}
\put(676,430){\rule[-0.200pt]{0.602pt}{1.400pt}}
\put(679,431){\rule[-0.200pt]{0.964pt}{1.400pt}}
\put(683,432){\rule[-0.200pt]{0.602pt}{1.400pt}}
\put(685,433){\rule[-0.200pt]{0.602pt}{1.400pt}}
\put(688,434){\rule[-0.200pt]{0.482pt}{1.400pt}}
\put(690,435){\rule[-0.200pt]{0.482pt}{1.400pt}}
\put(692,436){\rule[-0.200pt]{0.602pt}{1.400pt}}
\put(694,437){\rule[-0.200pt]{0.602pt}{1.400pt}}
\put(697,438){\rule[-0.200pt]{0.482pt}{1.400pt}}
\put(699,439){\rule[-0.200pt]{0.482pt}{1.400pt}}
\put(701,440){\rule[-0.200pt]{1.204pt}{1.400pt}}
\put(706,441){\rule[-0.200pt]{0.482pt}{1.400pt}}
\put(708,442){\rule[-0.200pt]{0.482pt}{1.400pt}}
\put(710,443){\rule[-0.200pt]{0.602pt}{1.400pt}}
\put(712,444){\rule[-0.200pt]{0.602pt}{1.400pt}}
\put(715,445){\rule[-0.200pt]{0.482pt}{1.400pt}}
\put(717,446){\rule[-0.200pt]{0.482pt}{1.400pt}}
\put(719,447){\rule[-0.200pt]{0.602pt}{1.400pt}}
\put(721,448){\rule[-0.200pt]{0.602pt}{1.400pt}}
\put(724,449){\rule[-0.200pt]{0.482pt}{1.400pt}}
\put(726,450){\rule[-0.200pt]{0.482pt}{1.400pt}}
\put(728,451){\rule[-0.200pt]{1.204pt}{1.400pt}}
\put(733,452){\rule[-0.200pt]{0.482pt}{1.400pt}}
\put(735,453){\rule[-0.200pt]{0.482pt}{1.400pt}}
\put(737,454){\rule[-0.200pt]{0.602pt}{1.400pt}}
\put(739,455){\rule[-0.200pt]{0.602pt}{1.400pt}}
\put(742,456){\rule[-0.200pt]{0.482pt}{1.400pt}}
\put(744,457){\rule[-0.200pt]{0.482pt}{1.400pt}}
\put(746,458){\rule[-0.200pt]{0.602pt}{1.400pt}}
\put(748,459){\rule[-0.200pt]{0.602pt}{1.400pt}}
\put(751,460){\rule[-0.200pt]{0.482pt}{1.400pt}}
\put(753,461){\rule[-0.200pt]{0.482pt}{1.400pt}}
\put(755,462){\rule[-0.200pt]{0.602pt}{1.400pt}}
\put(757,463){\rule[-0.200pt]{0.602pt}{1.400pt}}
\put(760,464){\rule[-0.200pt]{0.964pt}{1.400pt}}
\put(764,465){\rule[-0.200pt]{0.602pt}{1.400pt}}
\put(766,466){\rule[-0.200pt]{0.602pt}{1.400pt}}
\put(769,467){\rule[-0.200pt]{0.482pt}{1.400pt}}
\put(771,468){\rule[-0.200pt]{0.482pt}{1.400pt}}
\put(773,469){\rule[-0.200pt]{0.602pt}{1.400pt}}
\put(775,470){\rule[-0.200pt]{0.602pt}{1.400pt}}
\put(778,471){\rule[-0.200pt]{0.482pt}{1.400pt}}
\put(780,472){\rule[-0.200pt]{0.482pt}{1.400pt}}
\put(782,473){\rule[-0.200pt]{0.602pt}{1.400pt}}
\put(784,474){\rule[-0.200pt]{0.602pt}{1.400pt}}
\put(787,475){\rule[-0.200pt]{0.482pt}{1.400pt}}
\put(789,476){\rule[-0.200pt]{0.482pt}{1.400pt}}
\put(791,477){\rule[-0.200pt]{0.602pt}{1.400pt}}
\put(793,478){\rule[-0.200pt]{0.602pt}{1.400pt}}
\put(796,479){\rule[-0.200pt]{0.964pt}{1.400pt}}
\put(800,480){\rule[-0.200pt]{0.602pt}{1.400pt}}
\put(802,481){\rule[-0.200pt]{0.602pt}{1.400pt}}
\put(805,482){\rule[-0.200pt]{0.482pt}{1.400pt}}
\put(807,483){\rule[-0.200pt]{0.482pt}{1.400pt}}
\put(809,484){\rule[-0.200pt]{0.602pt}{1.400pt}}
\put(811,485){\rule[-0.200pt]{0.602pt}{1.400pt}}
\put(814,486){\rule[-0.200pt]{0.482pt}{1.400pt}}
\put(816,487){\rule[-0.200pt]{0.482pt}{1.400pt}}
\put(818,488){\rule[-0.200pt]{0.602pt}{1.400pt}}
\put(820,489){\rule[-0.200pt]{0.602pt}{1.400pt}}
\put(823,490){\rule[-0.200pt]{0.482pt}{1.400pt}}
\put(825,491){\rule[-0.200pt]{0.482pt}{1.400pt}}
\put(827,492){\rule[-0.200pt]{0.602pt}{1.400pt}}
\put(829,493){\rule[-0.200pt]{0.602pt}{1.400pt}}
\put(832,494){\rule[-0.200pt]{0.482pt}{1.400pt}}
\put(834,495){\rule[-0.200pt]{0.482pt}{1.400pt}}
\put(836,496){\rule[-0.200pt]{0.602pt}{1.400pt}}
\put(838,497){\rule[-0.200pt]{0.602pt}{1.400pt}}
\put(841,498){\rule[-0.200pt]{0.482pt}{1.400pt}}
\put(843,499){\rule[-0.200pt]{0.482pt}{1.400pt}}
\put(845,500){\rule[-0.200pt]{1.204pt}{1.400pt}}
\put(850,501){\rule[-0.200pt]{0.482pt}{1.400pt}}
\put(852,502){\rule[-0.200pt]{0.482pt}{1.400pt}}
\put(854,503){\rule[-0.200pt]{0.602pt}{1.400pt}}
\put(856,504){\rule[-0.200pt]{0.602pt}{1.400pt}}
\put(859,505){\rule[-0.200pt]{0.482pt}{1.400pt}}
\put(861,506){\rule[-0.200pt]{0.482pt}{1.400pt}}
\put(863,507){\rule[-0.200pt]{0.602pt}{1.400pt}}
\put(865,508){\rule[-0.200pt]{0.602pt}{1.400pt}}
\put(868,509){\rule[-0.200pt]{0.482pt}{1.400pt}}
\put(870,510){\rule[-0.200pt]{0.482pt}{1.400pt}}
\put(872,511){\rule[-0.200pt]{0.602pt}{1.400pt}}
\put(874,512){\rule[-0.200pt]{0.602pt}{1.400pt}}
\put(877,513){\rule[-0.200pt]{0.482pt}{1.400pt}}
\put(879,514){\rule[-0.200pt]{0.482pt}{1.400pt}}
\put(881,515){\rule[-0.200pt]{0.602pt}{1.400pt}}
\put(883,516){\rule[-0.200pt]{0.602pt}{1.400pt}}
\put(886,517){\rule[-0.200pt]{0.482pt}{1.400pt}}
\put(888,518){\rule[-0.200pt]{0.482pt}{1.400pt}}
\put(890,519){\rule[-0.200pt]{0.602pt}{1.400pt}}
\put(892,520){\rule[-0.200pt]{0.602pt}{1.400pt}}
\put(895,521){\rule[-0.200pt]{0.482pt}{1.400pt}}
\put(897,522){\rule[-0.200pt]{0.482pt}{1.400pt}}
\put(899,523){\rule[-0.200pt]{0.602pt}{1.400pt}}
\put(901,524){\rule[-0.200pt]{0.602pt}{1.400pt}}
\put(904,525){\rule[-0.200pt]{0.482pt}{1.400pt}}
\put(906,526){\rule[-0.200pt]{0.482pt}{1.400pt}}
\put(908,527){\rule[-0.200pt]{0.602pt}{1.400pt}}
\put(910,528){\rule[-0.200pt]{0.602pt}{1.400pt}}
\put(913,529){\rule[-0.200pt]{0.482pt}{1.400pt}}
\put(915,530){\rule[-0.200pt]{0.482pt}{1.400pt}}
\put(917,531){\rule[-0.200pt]{0.602pt}{1.400pt}}
\put(919,532){\rule[-0.200pt]{0.602pt}{1.400pt}}
\put(922,533){\rule[-0.200pt]{0.482pt}{1.400pt}}
\put(924,534){\rule[-0.200pt]{0.482pt}{1.400pt}}
\put(926,535){\rule[-0.200pt]{0.602pt}{1.400pt}}
\put(928,536){\rule[-0.200pt]{0.602pt}{1.400pt}}
\put(931,537){\rule[-0.200pt]{0.482pt}{1.400pt}}
\put(933,538){\rule[-0.200pt]{0.482pt}{1.400pt}}
\put(935,539){\rule[-0.200pt]{0.602pt}{1.400pt}}
\put(937,540){\rule[-0.200pt]{0.602pt}{1.400pt}}
\put(940,541){\rule[-0.200pt]{0.482pt}{1.400pt}}
\put(942,542){\rule[-0.200pt]{0.482pt}{1.400pt}}
\put(944,543){\rule[-0.200pt]{0.602pt}{1.400pt}}
\put(946,544){\rule[-0.200pt]{0.602pt}{1.400pt}}
\put(949,545){\rule[-0.200pt]{0.482pt}{1.400pt}}
\put(951,546){\rule[-0.200pt]{0.482pt}{1.400pt}}
\put(953,547){\rule[-0.200pt]{0.602pt}{1.400pt}}
\put(955,548){\rule[-0.200pt]{0.602pt}{1.400pt}}
\put(958,549){\rule[-0.200pt]{0.482pt}{1.400pt}}
\put(960,550){\rule[-0.200pt]{0.482pt}{1.400pt}}
\put(962,551){\rule[-0.200pt]{0.602pt}{1.400pt}}
\put(964,552){\rule[-0.200pt]{0.602pt}{1.400pt}}
\put(967,553){\rule[-0.200pt]{0.482pt}{1.400pt}}
\put(969,554){\rule[-0.200pt]{0.482pt}{1.400pt}}
\put(971,555){\rule[-0.200pt]{0.602pt}{1.400pt}}
\put(973,556){\rule[-0.200pt]{0.602pt}{1.400pt}}
\put(976,557){\rule[-0.200pt]{0.482pt}{1.400pt}}
\put(978,558){\rule[-0.200pt]{0.482pt}{1.400pt}}
\sbox{\plotpoint}{\rule[-0.500pt]{1.400pt}{1.400pt}}%
\put(264,133){\usebox{\plotpoint}}
\put(264,133){\rule[-0.500pt]{1.400pt}{29.631pt}}
\sbox{\plotpoint}{\rule[-1.000pt]{2.000pt}{2.000pt}}%
\put(264,256){\circle*{30}}
\end{picture}

\vspace*{-10mm}

\begin{center}
(a) \hspace{85mm} (b)
\end{center}

\begin{center}
{\small
\parbox{130mm}{
 {\bf Fig.~1:}
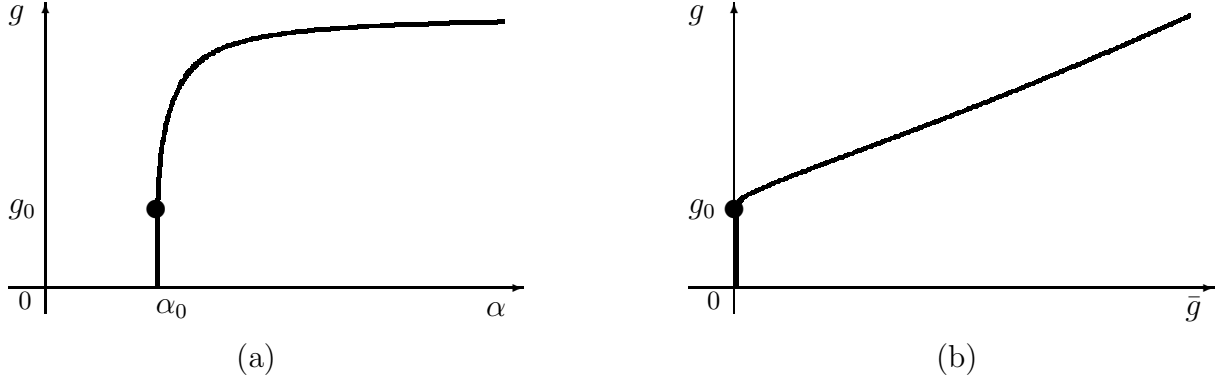
The phase diagram of the matrix model with perturbed polynomial potential.
It contains the phase of smooth surfaces $(g < g_0),$ intermediate phase
$(g=g_0)$ and phase of branched polymers $(g > g_0).$ (a): The dependence
of the critical value of cosmological constant on the touching constant;
(b): the effective touching constant along the line of criticality.
}}
\end{center}

\noindent
We found that the model has three phases with the string susceptibility
exponents $\gamma_{str}=-1/K$ for $g < g_0,$
$\gamma_{str}=1/(K+1)$ for $g = g_0$ and
$\gamma_{str}=1/2$ for $g > g_0.$ The position of the critical points in
shown in fig.~1(a). Let us consider the properties of the random surfaces
contributed at the partition function in different phases. The critical
surfaces have an infinite area and may touch each other. The average number
of touchings $\vev{T(\alpha)}=\frac1{N^2}\frac{dZ(\alpha)}{dg}$ we obtain
in the large $N$ limit using the definitions \re{2.1}, \re{2.9} and \re{2.10}
as
$$
\vev{T(\alpha)}=\frac{\alpha^2}{4}\VEV{\left(\frac1{N}\Tr M^2\right)^2}
       =\frac{\alpha^2}{4}\left(\VEV{\frac1{N}\Tr M^2}\right)^2 + \CO(N^{-2})
       =\frac{\geff^2}{4g^2} + \CO(N^{-2}).
$$
The dependence of the effective touching constant on $g$ is shown in
Fig.~1(b). Using it we find the average number of touchings as
$$
\vev{T(\alpha_{cr})}=0   \qquad   \mbox{for $g < g_0$}, \qqqquad
\vev{T(\alpha_{cr})} > 0 \qquad   \mbox{for $g < g_0.$}
$$
Thus, at the phase of multicritical Kazakov's points the
critical surfaces have not touchings and the perturbation becomes
irrelevant. It is interesting to compare this property with an analogous
one in the model \ci{Touching} corresponding to the solution
\re{7.4} for $K=2.$ By choosing $f(c)$ as a linear function of $c$ and
proceeding through relations \re{6.1}, \re{5.7} and \re{6.2} one gets that
in contrast with the previous case both the effective touching
constant $\geff$ and touching number $\vev{T(\alpha_{cr})}$ differ
from zero in the phase of smooth surfaces.

The behaviour of $\vev{T(\alpha)}$ near the critical line of fig.~1(a) can be
found using the asymptotics \re{7.10}, \re{7.12} and \re{7.14} of the
effective touching constant:
$$
\left.\vev{T(\alpha)}\right|_{g < g_c} \sim (\alpha-\alpha_{cr})^2
\qqqquad
\left.\vev{T(\alpha)}\right|_{g = g_c} \sim (\alpha-\alpha_{cr})^\frac{2K}{K+1}
\qqqquad
\left.\vev{T(\alpha)}\right|_{g > g_c} \sim (\alpha-\alpha_{cr})^\frac12
$$
These relations imply that under transition through the point $g=g_0$
the second derivative of the partition function
$
\frac1{N^2}\frac{d^2Z(\alpha)}{d\alpha dg}
=\frac{d\vev{T}}{d\alpha}\sim\frac{\geff}{2g^2}\frac{d\geff}{d\alpha}
$
undergoes an infinite jump.

There are two ways for critical random surfaces to increase the area:
to form an additional ``soap bubble'' or to increase of the area of the
existing ones. At the phases of smooth surfaces and branched polymers
only one of these mechanisms works. At the intermediate phase one
expects that both mechanisms are responsible for the formation of critical
surfaces.

\sect{Conclusion}

Let us summarize the properties of matrix models  perturbed by ``higher
order curvature term'' $(\Tr M^2)^2$ generating the touchings between
the random surfaces. We found that the matrix model with perturbed
{\it polynomial\/} potential has three phases: smooth surfaces, branched
polymers and intermediate phase. At the intermediate phase which appears for
special value $g_0$ of touching constant $g$ the string susceptibility
exponent has positive value $\gamma_{str}=1/(K+1)$ for some $K\geq 2.$
For $g<g_0$ the model turns into the phase of smooth surfaces corresponding
to $K-$th multicritical point. We obtained that the critical surfaces haven't
touchings. The perturbation becomes irrelevant in this phase and the model
is equivalent to $(2,2K-1)$ minimal conformal model coupled to gravity.
For $g>g_0$ the critical behaviour is dominated by the perturbation. The
random surfaces are degenerated into branched polymers and the string
susceptibility exponent approaches the maximum value $\gamma_{str}=1/2.$

In conclusion, we briefly discuss the critical behaviour of the perturbed
matrix model with Penner like potential defined in \re{2.7.1} and \re{5.2}.
As was shown in sect.~5.1, the
singularities of the function $f(c)$ at $c=-\eta^2$ play the central role in
an analysis of unperturbed model. The same is hold also for $g\neq 0.$
We have two sources of criticality in \re{4.2}: logarithmic peculiarities
of $\chi$ for
$c=-\eta^2$ and singularities of the derivatives $\frac{d\geff}{d\alpha}$
which define two different branches of possible critical points. At the
first case, the singularities of the functions $f(c)$ and $\phi(c)$ lead
in \re{4.3} to
$\alpha_{cr}=0$ for an arbitrary touching constant $g$ including
$g=0$ and string susceptibility has logarithmic behaviour \re{5.3}. At the
second case, the derivative \re{5.8} has singularities for the solutions of
\re{5.6}
and, moreover, there is a special point $c=-\eta^2$ at which the function
$\phi(c)$ becomes divergent. As was shown in sect.~5, for the
solutions of \re{5.6} the string susceptibility exponent is equal to
$\gamma_{str}=1/2.$
Being combined together, equations \re{5.6} and
\re{4.3} define the curve of criticality at the
$(\alpha, g)$ plane with the exponent $\gamma_{str}=\frac12.$
However, in despite of the model with polynomial
potential considered in the
previous sections this phase cannot be identified as the phase of
branched polymers because it has the following ``anomalous'' properties.
   Firstly, the cosmological constant
$\alpha_{cr}(g)$ decreases in this phase when the touching constant increases.
Recall, that at the phase of branched polymers, considered
in sect.~7, $\alpha_{cr}(g)$  has an opposite
behaviour shown in Fig.~1(a).
   Secondly, at the end point of this curve with $\alpha_{cr}=-2$ and
$g=4/\eta^4$ the parameter $c_{cr}$ calculated from \re{5.6} turns out to
be equal to $c_{cr}=-\eta^2$ giving rise to the singularities of
$f(c_{cr})$ and $\phi(c_{cr}).$ As a consequence, the derivative
$\frac{d\geff}{d\alpha}$
defined in \re{5.8} acquires additional singularities leading to the
increasing of the critical exponent at that point to $\gamma_{str}=2.$
This number lies outside the allowed region of values
$\gamma_{str}\leq 1/2$ one may get in any model of noninteracting planar
surfaces with positive weights \ci{Maximum}.
Moreover, the effective
touching constant scales near this point as
$\geff\sim 1/\sqrt{(\alpha-\alpha_{cr})}$ and
tends to infinity as $\alpha\to\alpha_{cr}.$
   Thirdly, the phase diagram of the model has two different
curves corresponding to $\gamma_{str}=0$ and $\gamma_{str}=2.$ It turns
out that despite of the analogous diagram for the perturbed matrix model with
polynomial potential shown in Fig.~1 these curves do not intersect each
other.
All these properties differs from what one expects to get by considering
the perturbed Penner model as discretization of a sum over surfaces for
Polyakov's bosonic string. This means that the Penner model originally
proposed to capture Euler characteristics of the moduli space of Riemann
surfaces fails to describe random surfaces after perturbation by higher
order curvature term.

\bigskip\bigskip
\noindent{\Large{\bf Note added}}
\bigskip

After the completion of this paper I was informed by L. Alvarez-Gaum\'e
that he got analogous results applying large $N$ reduced models to the
study of string with $c >1.$

\bigskip\bigskip
\noindent{\Large{\bf Acknowledgements}}
\bigskip

I am grateful to L.Alvarez-Gaum\'e, J.Ambj{\o}rn, S.Das, C.Destri,
G.Marchesini and E.Onofri for helpful discussions.

\newcommand\PL{Phys.~Lett.~{}}
\newcommand\NP{Nucl.~Phys.~{}}
\newcommand\PR{Phys.~Rev.~D}
\newcommand\CMP{Comm.~Math.~Phys.~{}}
\newcommand\AP{Ann.~Phys.~(NY)~{}}
\newcommand\RMP{Rev.~Mod.~Phys.~{}}
\newcommand\PTP{Prog.~Theor.~Phys.~{}}
\newcommand\CQG{Class.~Quantum~Grav.~{}}
\newcommand\MPL{Mod.~Phys.~Lett.~{}}
\newcommand\PRept{Phys.~Rept.~{}}
\newcommand\PRL{Phys.~Rev.~Lett.~{}}

\bb{99}
\bi{KPZ}
         A.M.Polyakov, \MPL A2 (1987) 893;
\\       V.G.Knizhnik, A.M.Polyakov and A.B.Zamolodchikov, \MPL A3 (1988) 819.
\bi{Num}
         F.David, \NP B257 (1985) 45;
\\       J.Ambj{\o}rn, B.Durhuus and J.Fr{\"o}hlich, \NP B257 (1985) 433;
\\       D.V.Boulatov, V.A.Kazakov, I.K.Kostov and A.A.Migdal,
         \NP B257 (1985) 641; \PL 174B (1986) 87.
\bi{Pol}
         A.Polyakov, {\it ``Singular states in $2D$ quantum gravity''},
         PUPT-1289 (September, 1991).
\bi{Touching}
         S.R.Das, A.Dhar, A.M.Sengupta and S.R.Wadia, \MPL A5 (1990) 1041.
\bi{g>0}
         C.Marzban abd R.R.Viswanathan, \PL 277B (1992) 289;
\\       E.Br{\'e}zin and S.Hikami, {\it ``A naive matrix-model approach to
         two-dimensional quantum gravity coupled to matter of arbitrary central
         charge''}, LPTENS 92/10 (March, 1992).
\bi{Review}
         c.f. review L.Alvarez-Gaum\'e, Helv.~Phys.~Acta 64 (1991) 359.
\bi{Multi}
         V.A.Kazakov, \MPL A4 (1989) 2125.
\bi{Penner1}
         J.Distler and C.Vafa, \MPL A6 (1991) 259.
\bi{Penner2}
         C.-I.Tan, \MPL A6 (1991) 1373;
\\       S.Chaudhuri, H.Dykstra and J.Lykken, \MPL A6 (1991) 1665.
\bi{MinCFT}
         D.Gross and A.Migdal, \PRL 64 (1990) 717;
\\       E.Br{\'e}zin, M.Douglas, V.Kazakov and S.Shenker, \PL 237B (1990) 43;
\\       \v{C}.Crnkovi\'c, P.Ginsparg and G.Moore, \PL 237B (1990) 196;
\\       M.Staudacher, \NP B336 (1990) 349.
\bi{GenPen}
         L.Chekhov and Yu.Makeenko, \PL 278B (1992) 271;
         {\it ``The multicritical Kontse\-vich\---Penner model''}, NBI-HE-92-03
         (January, 1992);
\\       J.Ambj{\o}rn, L.Chekhov and Yu.Makeenko, {\it ``Higher genus
         correlators and $W-$infinity from the hermitian one-matrix
         model''}, NBI-HE-92-22 (February, 1992).
\bi{Loop1}
         A.A.Migdal, \PRept 102 (1983) 199.
\bi{Loop2}
         F.David, \MPL A5 (1990) 1019.
\bi{Loop3}
         Yu.Makeenko, \MPL A6 (1991) 1901.
\bi{Maximum}
         J.Ambj{\o}rn, B.Durhuus, J.Fr{\"o}hlich and P.Orland, \NP B270
         (1986) 457;
\\       J.Ambj{\o}rn, B.Durhuus and J.Fr{\"o}hlich, \NP B275 (1986) 161;
\eb
\end{document}